\documentclass[aps,pra,longbibliography,reprint,amsmath,amssymb,superscriptaddress]
{revtex4-2}
\usepackage{graphicx}
\usepackage{dcolumn}
\usepackage{bm}

\usepackage{xcolor}
\usepackage{soul}
\usepackage{graphicx}
\usepackage{dcolumn}
\usepackage{bm}
\usepackage{mathrsfs}
\usepackage{hyperref}
\usepackage{amsmath}
\def\lesssim{\ \raise.3ex\hbox{$<$}\kern-0.8em\lower.7ex\hbox{$\sim$}\ }
\def\gesim{\ \raise.3ex\hbox{$>$}\kern-0.8em\lower.7ex\hbox{$\sim$}\ }

\def\ev#1{\langle #1 \rangle}

\def\up{\uparrow}
\def\down{\downarrow}

\begin{document}

\title{Two-body loss dynamics as a probe of dynamical bulk viscosity \\ in ultracold atomic gases}

\author{Tingyu Zhang}
\affiliation{Department of Physics, State Key Laboratory of Optical Quantum Materials, and Hong Kong Institute of Quantum Science and Technology, University of Hong Kong, Hong Kong, China}
\author{Hiroyuki Tajima}
\affiliation{Department of Physics, Graduate School of Science, The University of Tokyo, Tokyo 113-0033, Japan}
\affiliation{RIKEN Nishina Center, Wako 351-0198, Japan}
\author{Takeo Kato}
\affiliation{Institute for Solid State Physics, The University of Tokyo, 5-1-5 Kashiwanoha, Kashiwa 277-8581, Japan}

\begin{abstract}
We show that the equilibrium connected correlation of the two-body loss rate provides access to the dynamical bulk viscosity in weakly dissipative quantum gases. Starting from the Lindblad equation for weak inelastic losses, we derive the loss-current operator and, using a counting-field formulation of the loss statistics, the relation between the statistics of detected losses and the equilibrium loss-rate correlator. The connected correlation of the two-body loss rate is found to be proportional to the equilibrium contact correlation and hence to the bulk-viscosity spectrum. We show that the measured loss-counting covariance contains, besides the Poissonian self-correlation, a quantum-jump backaction term of the same order as the target correlator, which is fixed by the transient of the mean loss rate after the loss is switched on. The bulk-viscosity spectrum can therefore be reconstructed from the combination of the loss-counting covariance and the mean loss dynamics. 
Our result identifies weak two-body loss dynamics as a probe of dynamical bulk viscosity, whose measurement has remained elusive in experiments.

\end{abstract}

\maketitle

\section{Introduction}
Nonequilibrium phenomena and transport coefficients play a pivotal role in modern physics, as they reveal how interacting quantum systems respond to external perturbations and relax toward equilibrium. Their importance extends across a wide range of fields, from condensed-matter to high-energy physics. In particular, phenomena such as collective-mode damping~\cite{PhysRevA.72.043605,Punk_2006}, diffusive transport~\cite{doi:10.1126/science.1247425,PhysRevLett.118.130405,PhysRevA.85.013636,PhysRevA.86.013617}, and hydrodynamic expansion~\cite{doi:10.1126/science.1079107,doi:10.1126/science.1195219,PhysRevLett.106.115304} are governed by transport coefficients, which therefore provide essential information about correlation effects and nonequilibrium dynamics beyond static thermodynamics. The study of transport has become especially prominent in strongly interacting systems, where conventional quasiparticle descriptions may fail and hydrodynamics emerges as a universal low-energy framework.

The discovery of $^4$He superfluid triggered the study of quantum hydrodynamics and transport, and
contributed to the modern concept of ``perfect fluidity" in strongly interacting quantum matter.
Such fluids have extremely small shear viscosity and their measurements led to important paradigms including high-$T_{\rm c}$ superconductors~\cite{bednorz1986possible,PhysRevB.90.134509} and quark gluon plasma~\cite{yagi2005quark}.
In this context, the dynamical bulk viscosity is of particular interest because it describes dissipation under isotropic compression or expansion~\cite{Landau,forster2018hydrodynamic,PhysRevA.81.053610}. 
The bulk viscosity is also relevant for understanding neutron-star matter~\cite{alford2007bulk,PhysRevD.39.3804,PhysRevD.105.103016,PhysRevD.110.L061303,sedrakian2026nuclear}.
Microscopically, the bulk viscosity reflects the finite relaxation time required for the internal degrees of freedom of the fluid to re-equilibrate after a density perturbation, and is associated with the breaking of scale invariance~\cite{PhysRevLett.108.070404}. The bulk viscosity vanishes in scale-invariant systems such as unitary gases~\cite{PhysRevLett.98.020604,ENSS2011770}, and becomes finite when this symmetry is broken, for example by interactions~\cite{PhysRevLett.111.120603,PhysRevA.81.053610}, finite-range effects~\cite{PhysRevLett.125.240402}, or the presence of bound states~\cite{PhysRevLett.118.130405,FUJII2023169296}. Therefore, the bulk viscosity is of particular interest in strongly correlated systems such as ultracold atomic gases in the crossover between the Bardeen-Cooper-Schrieffer (BCS) and Bose-Einstein-condensate (BEC) regimes, where it encodes nontrivial information about the nonequilibrium dynamics beyond ideal hydrodynamics. However, its direct measurement remains elusive~\cite{PhysRevA.96.063607}.

\begin{figure}[t]
    \centering
    \includegraphics[width=\linewidth]{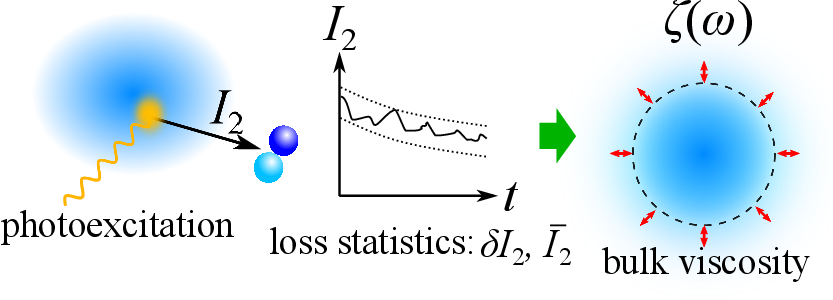}
    \caption{
    Schematic of the proposal for measuring the dynamical bulk viscosity $\zeta(\omega)$ from the statistics (fluctuations and mean transient) of the photoexcited two-body loss current $I_2$.}
    \label{fig:1}
\end{figure}

On the other hand, atom-loss measurements have become a standard diagnostic tool in ultracold-atom experiments~\cite{RevModPhys.82.1225,PhysRevLett.100.053201}. A recent measurement based on photoexcitation of atom pairs has reached sufficient precision to extract Tan's contact across the BCS-BEC crossover~\cite{PhysRevLett.132.263401}, where the two-body loss rate is directly linked to the total contact. This inspires us to connect the bulk viscosity with the fluctuation of loss current, since the former can be written as a correlation function of the contact~\cite{PhysRevLett.123.205301,PhysRevA.102.023310} and the latter is defined as the current-current correlation function~\cite{BLANTER20001}. Moreover, the loss rate can be tuned smoothly from BCS to BEC limit, providing a high controllability of loss strength in cold atomic systems, due to the tunable interparticle scattering length.

In this work, we propose a measurement method for the bulk viscosity based on the statistics of atom loss, namely the loss-counting covariance combined with the transient of the mean loss rate, in a two-component Fermi gas that remains in a quasi-stationary thermal state (see Fig.~\ref{fig:1}). We start from a multi-channel loss model including both one-body and two-body channels and show that the equilibrium connected correlation of the two-body loss rate is proportional to the dynamical bulk-viscosity spectrum. 
We then derive, from the number-resolved Lindblad equation, how this correlator is related to the covariance of detected losses: in addition to the leading Poissonian shot noise, the covariance contains a quantum-jump backaction term of the same order, which is determined by the transient mean loss rate and can be subtracted.

The Fano factor, defined by the noise-to-current ratio, can serve as a probe of effective charge carried by the nonequilibrium current, and thus identify multi-body processes from one-body ones~\cite{PhysRevLett.79.2526,10.1093/pnasnexus/pgad045,PhysRevApplied.21.L031001,10.1093/pnasnexus/pgad045,PhysRevB.111.054502}.
This idea has also been applied to non-equilibrium Lindblad dynamics of spin fluctuations excited by pulsed light~\cite{PhysRevLett.134.106702}.
Thus the dominant two-body loss can be ensured if the value of the total Fano factor is close to that of a pure two-body loss. This study establishes a realistic measurement proposal for bulk viscosity in quantum gases by monitoring
the fluctuations and the mean dynamics of two-body atom loss under weak dissipation.

The outline of this paper is as follows: In Sec.~\ref{Sec2}, we present the theoretical model for a weakly dissipative Fermi gas with two loss channels based on the Lindblad equation and an equilibrium Hamiltonian. In Sec.~\ref{Sec3}, we derive the formulas of the noise power spectrum of two-body loss and the bulk viscosity spectrum based on the second-order quantum virial expansion. Finally we summarize this paper and give perspectives in Sec.~\ref{Sec4}.

\section{Multi-channel loss model}\label{Sec2}

Throughout the paper, we take $\hbar = k_B =1$.
We consider a two-component Fermi gas with weak one- and two-body losses. The system is described by a Lindblad master equation
\begin{align}\label{eq:lindblad}
\frac{d \rho}{d t}=-i[H,\rho]+&\sum_n\int d^3\bm r\Big[L_n(\bm r)\rho L_n^\dagger(\bm r)\nonumber\\
&-\frac{1}{2}\left\{L_n^\dagger(\bm r)L_n(\bm r),\rho\right\}\Big],
\end{align}
where $\rho$ is the total density matrix, $L_n$ are Lindblad operators for one-body ($n=1\uparrow, 1\downarrow$ with pseudospin $\sigma=\up,\down$) and two-body ($n=2$) loss processes 
\begin{align}
\label{eq:2}
L_{1\sigma}(\bm r)&=
\sqrt{\gamma_{1\sigma}}\,\psi_\sigma(\bm r),\nonumber\\
L_2(\bm r)&=\sqrt{\gamma_2}\,\psi_{\down}(\bm r)\psi_{\up}(\bm r).
\end{align}
Here $\gamma_{1\sigma}$ and $\gamma_2$ respectively denote the one-body and two-body loss rates, and $\psi_\sigma(\bm{r})$ is the field operator of $\sigma$ component. The loss current operator is defined by 
\begin{align}
    I=&-\frac{d\langle N\rangle}{dt}\nonumber\\
    =&\frac{1}{2}
    \sum_{n}\int d^3\bm{r}[N L_n^\dagger(\bm{r})L_n(\bm{r})
    + L_n^\dagger(\bm{r})L_n(\bm{r}) N\nonumber\\
    &-2 L_n^\dagger (\bm{r})N L_n(\bm{r})],
\end{align}
where $N=\sum_\sigma\int d^3\bm r\,\psi_\sigma^\dagger(\bm r)\psi_\sigma(\bm r)$ is the total particle number operator, and only the dissipative jump processes contribute to the current operator. We notice that $[L_{1\sigma},N]=L_{1\sigma}$ and $[L_2,N]=2L_2$, meaning the one-body jump operator removes a single atom for each jump, while the two-body jump operator removes a pair. Then the loss current operator can be rewritten as
\begin{align}\label{eq:current}
I=\int d^3\bm r\left[\sum_\sigma \gamma_{1\sigma}\psi_\sigma^\dagger\psi_\sigma+2\gamma_2\psi_\up^\dagger\psi_\down^\dagger\psi_\down\psi_\up\right].
\end{align}
where the first and second term respectively correspond to the one-body and two-body loss current and it can be rewritten as $I=I_1+I_2$.

We focus on the weak-loss regime, where $N/\langle I\rangle=\tau_{\rm loss} \gg  \tau_c$ with $\tau_{\rm loss}$ and $\tau_c$ respectively denoting the time scale of particle loss and the intrinsic correlation time. The measurement window $\tau$ is given by $\tau_c \ll \tau \ll \tau_{\rm loss}$. In this window the contact correlations have decayed while only a negligible fraction of atoms is removed, so that the density, temperature, and equilibrium state change only weakly, namely, the system remains in a quasi-steady thermal state during the noise measurement.
We therefore introduce the quasi-stationary zero-frequency noise power~\cite{blanter2000shot,RevModPhys.81.1665}
\begin{align}
    S_N=\lim_{ \tau_c\ll\tau\ll\tau_{\rm loss}}\frac{\langle [N(\tau)-N(0)]^2\rangle-\langle [N(\tau)-N(0)]\rangle^2 }{\tau},
\end{align}
where $N(t)$ is the Heisenberg representation of the particle number operator.
In weak-dissipation limit, the noise is expanded as
\begin{align}
\label{eq:sn}
    S_N&=
    \int d^3\bm{r}\sum_{\sigma}\left\langle\gamma_{1\sigma}\psi_\sigma^\dag\psi_\sigma
    +2\gamma_2\psi_\up^\dag \psi_\down^\dag \psi_\down\psi_\up
    \right\rangle\cr
    &+\int_0^{\infty}dt \big[\langle \{\delta I(t),\delta I(0)\}\rangle+\mathcal{B}(t)\big]+O(\gamma_{1,2}^3),
\end{align}
with $\delta I(t)=I(t)-\langle I(t) \rangle$.
Here the first term is the Poissonian shot-noise contribution associated with individual one- and two-body loss processes, and thus can be used to identify the elementary carrier. The term $\mathcal{B}(t)$ represents the quantum-jump backaction correction~\cite{PhysRevLett.134.106702}, arising because the first detected loss event changes the system state and thereby modifies the probability of subsequent loss events.
Meanwhile, we are interested in the next leading-order terms associated with the transport coefficient.
The connected current-current correlator can be evaluated with the equilibrium Hamiltonian
\begin{align}
H=\int d^3&\bm r\sum_\sigma \psi_\sigma^\dagger(\bm r)\left(-\frac{\nabla^2}{2m}\right)\psi_\sigma(\bm r)\nonumber\\
&+g\int d^3\bm r\,\psi_\up^\dagger(\bm r)\psi_\down^\dagger(\bm r)\psi_\down(\bm r)\psi_\up(\bm r)\nonumber\\
\equiv H_0+&U,
\end{align}
where $g$ is the two-body interaction strength. We define the loss-noise correlator as the two-time connected correlation function of the loss current,
\begin{align}
\mathcal D(t)=\frac{1}{2}\ev{\{\delta I(t),\delta I(0)\}},
\end{align}
where $I(t)=e^{iHt} Ie^{-iHt}$ denotes the time-dependent loss current operator, and $\rho_0$ is the density matrix of a thermal
equilibrium state. Notice that the current operator can be rewritten as 
\begin{align}
    I = \sum_{\sigma}\gamma_{1\sigma} N_{\sigma}+\frac{2\gamma_2}{g}U.
\end{align}
Then the connected correlator can be rewritten as 
\begin{align}
    \mathcal{D}(t)=&\frac{1}{2}\Bigg\langle\Bigg\{\sum_{\sigma}\gamma_{1\sigma}\delta N_{\sigma}(t)+\frac{2\gamma_2}{g}\delta U(t)\nonumber,\\
    &\sum_{\sigma'}\gamma_{1\sigma'}\delta N_{\sigma'}(0)+\frac{2\gamma_2}{g}\delta U(0)\Bigg\}
    \Bigg\rangle
\end{align}
where $\delta U(t)=U(t)-\langle U(t)\rangle$, and hereafter $\langle \cdots\rangle$ denotes the thermal average in the absence of dissipation.

For the only one-body loss channel, we find
\begin{align}
    \mathcal{D}_{1}(t)&=\frac{1}{2}\langle \{\delta I_1(t),\delta I_1(0)\}\rangle\cr
    &=\sum_{\sigma,\sigma'}
    \gamma_{1\sigma}\gamma_{1\sigma'}\frac{1}{2}
    \langle \{\delta N_{\sigma}(t),\delta N_{\sigma'}(0)\}\rangle.
\end{align}
Notice that $N_{\sigma}$ is conserved by the equilibrium Hamiltonian since $[H,N_{\sigma}]=0$. Therefore $N_{\sigma}(t)=N_{\sigma}(0)$ and $\mathcal{D}_{1}$ probes the static number fluctuations. In particular, for spin-independent one-body loss where $\gamma_{1\uparrow}=\gamma_{1\downarrow }\rightarrow\gamma_1$ and $N_{\uparrow}=N_{\downarrow}=N/2$, $\mathcal{D}_{1}$ becomes
\begin{align}
    \mathcal{D}_{1}(t)=
    \gamma_1^2\langle (\delta N)^2\rangle,
\end{align}
which is associated with the isothermal compressibility characterizing how sensitively the particle number responds to the chemical potential: 
\begin{align}
    \kappa =\frac{V}{\langle N\rangle^2}\left(\frac{\partial \langle N\rangle}{\partial\mu}\right)_T.
\end{align}
The relation between one-body loss fluctuation and the compressibility can be given by (see Appendix~\ref{appendixA})
\begin{align}\label{D1-kappa}
    \mathcal{D}_{1}=\gamma^2_1\frac{T\langle N\rangle^2}{V}\kappa.
\end{align}

Indeed, in ultracold Fermi gases, the compressibility has been experimentally measured by analyzing the speckle pattern of scattered light~\cite{PhysRevLett.105.040402,PhysRevLett.106.010402}. This part, however, is purely static and does not contain dynamical information.

\section{Contact correlator and bulk viscosity}\label{Sec3}

The nontrivial dynamical content arises from the two-body channel. To make its connection to contact correlations explicit, we introduce the contact density operator
\begin{align}\label{eq:contact-def}
C(\bm r)=m^2g^2\psi_\up^\dagger(\bm r)\psi_\down^\dagger(\bm r)\psi_\down(\bm r)\psi_\up(\bm r),
\end{align}
which is proportional to the local pair density.
The current operator for the two-body loss can be written as
\begin{align}
    I_2&=2\gamma_2\int d^3\bm{r}\,\psi_{\up}^\dag(\bm{r})\psi_{\down}^\dag(\bm{r})\psi_{\down}(\bm{r})\psi_{\up}(\bm{r})\cr
    &\equiv \frac{2\gamma_2}{m^2g^2}\int d^3\bm{r}\,C(\bm{r}).
\end{align}
In the weak loss limit its expectation value can be given by $\langle I_2\rangle =2\gamma_2VC_0/(m^2g^2)$, where $V$ is the volume of the system and $C_0$ is Tan's contact density in a homogeneous system. Recently, the thermal average of the contact density $C(\bm{r})$ has been measured experimentally from the two-body loss current in a Fermi gas~\cite{PhysRevLett.132.263401} as well as in a Bose gas~\cite{prf8-3q27}.

In the weak loss regime, the connected correlator associated with the two-body loss channel reads
\begin{align}
    \mathcal{D}_{2}(t)
    &=\frac{1}{2}\langle\{\delta I_2(t),\delta I_2(0)\}\rangle\cr
    &=\left(\frac{2\gamma_2}{m^2g^2}\right)^2\int d^3\bm{r}\int d^3\bm{r}'\frac{1}{2}
    \langle \{\delta C(\bm{r},t),\delta C(\bm{r}',0)\}\rangle\nonumber\\
    &=\left(\frac{2\gamma_2}{m^2g^2}\right)^2
    V\int d^3\bm{r}\frac{1}{2}
    \langle \{\delta C(\bm{r},t),\delta C(\bm{0},0)\}\rangle,
\end{align}
where the fluctuation reads $\delta C(\bm{r},t)=C(\bm{r},t)-C_0$, and $\int d^3\bm{r}\langle\{\delta C(\bm{r},t),\delta C(\bm{r}',0)\}\rangle=\int d^3\bm{r}\langle \{\delta C(\bm{r},t),\delta C(\bm{0},0)\}\rangle$ is guaranteed by translational invariance. The corresponding frequency-resolved noise spectrum is obtained from the Fourier transform
\begin{align}\label{eq:Domega-Fourier}
    \mathcal{D}_{2}(\omega)=\int_{-\infty}^{\infty} dt \,e^{i\omega t} \mathcal{D}_{2}(t).
\end{align}
The second bridge comes from the Kubo formula for the bulk viscosity. In a scale-anomalous Fermi gas, the dynamical bulk viscosity is controlled by the retarded contact correlator~\cite{PhysRevA.102.023310},
\begin{align}\label{eq:kubo-zeta}
\zeta(\omega)=\frac{c_d^2}{\omega d^2}{\rm Im}[G_{CC}^{\rm R}(\omega)]
\end{align}
where 
\begin{align}\label{eq:retarded-correlator}
    G_{CC}^{\rm R}(\omega)=i
    \int_0^{\infty}dt \, e^{i(\omega+i0^+) t}
    \int d^d\bm{r}\, \langle[C(\bm{r},t),C(\bm{0},0)]\rangle 
\end{align}
is the retarded contact correlator,
$c_d=-\partial g^{-1}/\partial\ln|a|\,/m^2$ is the beta function~\cite{PhysRevLett.123.205301}, and $a$ is the $s$-wave scattering length. Particularly at $d=3$, $\frac{m}{4\pi a}={g}^{-1}+\sum_{\bm{p}}\frac{m}{p^2}$ and we have
\begin{align}
    c_{d=3}=-\frac{1}{m^2}\frac{\partial}{\partial\ln|a|}\left(\frac{m}{4\pi a}\right)=\frac{1}{4\pi m a}
\end{align}
We introduce the momentum cutoff $\Lambda$ for the s-wave interaction strength and the effective range $r_{\rm eff}=4/(\pi \Lambda)$. Then the interaction strength $g$ can be rewritten as 
\begin{align}
    g=\bigg[\frac{m}{4\pi a}-\frac{2m}{\pi^3 r_{\rm eff}}\bigg]^{-1}.
\end{align}
For the dissipative system with an inelastic two-body loss,
we then introduce a complex scattering length, defined by~\cite{PhysRevA.95.012708,PhysRevA.103.013724}
\begin{align}
    \frac{m}{4\pi(a-ia_I)}&=\frac{1}{g-i\gamma_2/2}+\sum_{\bm{p}}\frac{m}{p^2}.
\end{align}
The imaginary part is given by 
\begin{align}
    \frac{ma_I}{4\pi(a^2+a_I^2)}=\frac{\gamma_2/2}{g^2+\gamma_2^2/4}.
\end{align}
In weak loss limit assuming $|a_I|\ll|a|$ and $\gamma_2\ll g$, we obtain the imaginary scattering length
\begin{align}\label{eq:aI}
    a_I\simeq
    \bigg({1}-\frac{8a}{\pi^2 r_{\rm eff}}\bigg)^2
    \frac{m\gamma_2}{8\pi}.
\end{align}

To connect the measurable loss-noise spectrum to a transport coefficient, we use the fluctuation-dissipation theorem together with the Kubo-Martin-Schwinger (KMS) condition~\cite{Kubo_1966,mahan2013many}, which relates the spontaneous fluctuations of an observable to the dissipative part of its linear response function. Notice that the commutation in Eq.~(\ref{eq:retarded-correlator}) $[C(\bm{r},t),C(\bm{0},0)]=[\delta C(\bm{r},t),\delta C(\bm{0},0)]$
we obtain that
\begin{align}
\label{eq:19rev}
    {\rm Im}G_{CC}^{\rm R}(\omega)=\frac{1-e^{-\beta\omega}}{2}G_{CC,c}^{>}(\omega),
\end{align}
where
\begin{align}
\label{eq:20rev}
    G_{CC,c}^{>}(\omega)&=\int_{-\infty}^{\infty}dt e^{i\omega t}
    \int d^d\bm{r}
    \,\langle \delta C(\bm{r},t)\delta C(\bm{0},0)\rangle\cr
    &=\frac{1}{V}\frac{2}{1+e^{-\beta\omega}}\left(\frac{m^2g^2}{2\gamma_2}\right)^2\mathcal{D}_{2}(\omega)
\end{align}
is the greater connected contact correlator.
Combining the Kubo formula with the fluctuation-dissipation relation, we obtain a connection between the connected correlation of two-body loss and the bulk-viscosity spectrum,
\begin{align}\label{eq:main-relation}
    \mathcal{D}_{2}(\omega)=
        \left(\frac{2\gamma_2}{m^2g^2}\right)^2
    V
    \coth{\left(\frac{\beta\omega}{2}\right)}\frac{\omega d^2}{c_d^2}\zeta(\omega),
\end{align}
showing that the frequency-resolved correlation spectrum of the two-body loss current is proportional to the dynamical bulk viscosity, up to known prefactors.

\begin{figure}[t]
    \includegraphics[width=8cm]{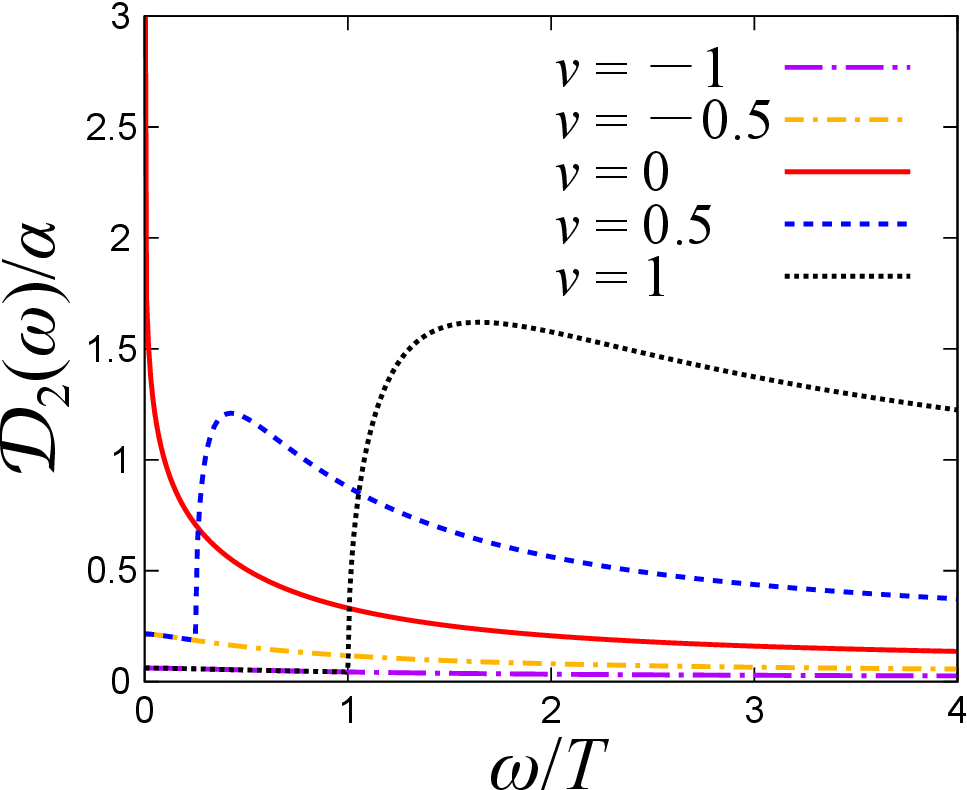}
    \caption{Dimensionless connected correlation spectrum of the two-body loss current obtained from the second-order virial expansion with different values of $v$.}
    \label{fig:noise}
\end{figure}
To obtain the analytical result of the bulk viscosity, we use quantum virial expansion for a homogeneous three-dimensional Fermi gas~\cite{FUJII2023169296}. This method is well suited to the high-temperature regime, where the fugacity $z=e^{\beta\mu}$ is small and thermodynamic quantities can be expanded systematically in powers of $z$~\cite{PhysRevLett.111.120603,PhysRevA.88.043636}.
Writing $v=\lambda/(a\sqrt{2\pi})$ and $\lambda=\sqrt{2\pi/(mT)}$, the second order expansion gives
\begin{align}\label{eq:virial-zeta}
\frac{d^2}{c_d^2}\zeta(\omega)
=&32\sqrt{2}\pi^2m z^2\lambda^{-3}\frac{1-e^{-\beta\omega}}{\omega}\nonumber\\
&\Biggl[
2ve^{v^2}\theta(v)
\frac{\sqrt{\beta(\omega-E_b)}\,\theta(\omega-E_b)}{\beta\omega}
\nonumber\\
&
+\frac{1}{\pi}\int_0^{\infty}d y\,
\frac{e^{-y}\sqrt{y(y+\beta\omega)}}{(y+v^2)(y+\beta\omega+v^2)}
\Biggr],
\end{align}
where $E_b=1/(ma^2)$ is the two-body bound-state binding energy. The factor $\theta(v)$ indicates that the first term exists only when $v>0$, namely $a>0$, where a two-body bound state is permitted. Substituting Eq.~(\ref{eq:virial-zeta}) into Eq.~(\ref{eq:main-relation}) gives the two-body loss correlation spectrum,
\begin{align}\label{eq:virial-noise}
\mathcal {D}_2(\omega)=&\alpha\frac{1+e^{-\beta\omega}}{2}
\Biggl[
2ve^{v^2}\theta(v)
\frac{\sqrt{\beta(\omega-E_b)}\,\theta(\omega-E_b)}{\beta\omega}\nonumber\\
&+\frac{1}{\pi}\int_0^{\infty}d y\,
\frac{e^{-y}\sqrt{y(y+\beta\omega)}}{(y+v^2)(y+\beta\omega+v^2)}
\Biggr],
\end{align}
with
\begin{align}
\alpha=&\left(\frac{2\gamma_2}{m^2g^2}\right)^2V\,64\sqrt{2}\pi^2m z^2\lambda^{-3}\nonumber\\
\equiv&\bigg(\frac{a_I}{a^2}\bigg)^2 V 64\sqrt{2}m^{-1}z^2\lambda^{-3}.
\end{align}
The first term of Eq.~(\ref{eq:virial-noise}) comes from the contribution of bound states, while the second term corresponds to a scattering continuum of unbound two-particle states.
Especially at unitarity, the scattering length diverges and $v\rightarrow0$, $\mathcal{D}_2(\omega)$ is given by
\begin{align}\label{eq:D-unitary}
    \mathcal{D}_2(\omega)=\frac{\alpha(1+e^{-\beta\omega})}{\pi}
    \int_0^{\infty}ds\frac{e^{-s^2}}{\sqrt{s^2+\beta\omega}},
\end{align}
where we take $y=s^2$ and $dy=2sds$ to avoid the singularity at $y=0$.

Fig.~\ref{fig:noise} shows the noise power spectrum $\mathcal{D}(\omega)$ with different values of $v$, involving positive and negative $a$. 
According to Eq.~(\ref{eq:virial-noise}) and Eq.~(\ref{eq:D-unitary}), we can see that at large $\omega$, the correlation spectrum decays as 
\begin{align}
    \mathcal{D}_2(\omega)\propto\frac{1}{\sqrt{\beta\omega}}\ \ , \quad (\omega\rightarrow\infty).
\end{align}
According to Eq.~(\ref{eq:main-relation}), we can consequently obtain the frequency dependence of the bulk viscosity as
\begin{align}\label{eq:viscosity-noise}
    \zeta(\omega)
    =
    \frac{1}{V}\tanh{\left(\frac{\beta\omega}{2}\right)}\frac{c_d^2}{\omega d^2}
    \left(\frac{m^2g^2}{2\gamma_2}\right)^2
    \mathcal{D}_2(\omega).
\end{align}
The two-body loss rate $\gamma_2$ linked to the contact density can be experimentally measured based on high-field absorption imaging~\cite{PhysRevLett.132.263401}, where the loss is induced through photoexcitation of diatomic pairs, and the slow decay of the total atom number ensures that the system stays in thermal equilibrium state. This approach for contact measurement is in sharp contrast to the previous measurements where the shear viscosity in the DC limit was estimated from the hydrodynamic flow directly~\cite{cao2011universal,PhysRevLett.113.020406,PhysRevLett.128.090402}.
In addition, the contact dynamics has also been probed in Bose gases through a two-body loss caused by photodissociation of closed-channel molecules~\cite{prf8-3q27}, where the closed channel is quenched to Feshbach resonance using rapid optical control.
\begin{figure}
    \centering
    \includegraphics[width=0.9\linewidth]{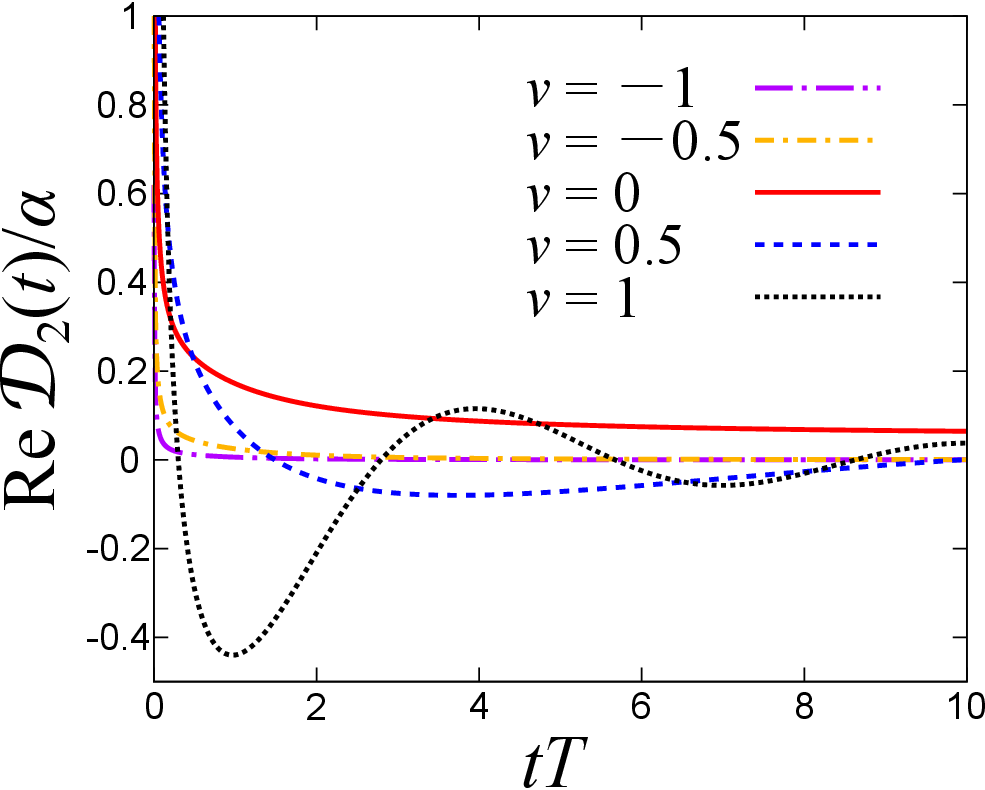}
    \caption{Time dependence of noise ${\rm Re}\mathcal{D}_2(t)$ at different $v$, where the smooth frequency cutoff $w(\omega)=e^{-(\omega/\omega_{\rm c})^2}$ with $\omega_{\rm c}=300T$ is used practically in Eq.~\eqref{eq:redt}. While $S_N$ in Eq.~\eqref{eq:sn} is always positive, ${\rm Re}\mathcal{D}_2(t)$ can be negative as found in the case with $v=0.5,1$ due to the oscillation of the bound-state contribution. At $t=0$, $\mathcal{D}_2(t)$ diverges due to $\int_0^{\infty}d\omega \frac{1}{\sqrt{\beta\omega}}\cos(\omega t)\propto1/\sqrt{t}$.}
    \label{fig:noise_tdep}
\end{figure}

We now discuss how the equilibrium two-body loss-rate correlator $\mathcal{D}_2(t)$ can be reconstructed from experimentally detected two-body loss and its noise. We consider the experimentally detected atom-loss record and derive its covariance from the underlying quantum-jump dynamics. For pure two-body loss, the exact counting statistics are obtained from the number-resolved Lindblad equation, as shown in Appendix~\ref{appendixB}. We define
\begin{align}\label{eq:loss-record}
    J(t)=\sum_{\sigma,i}\delta\big(t-t^{(1)}_{i,\sigma}\big)+2\sum_{j}\delta\big(t-t_j^{(2)}\big),
\end{align}
denoting the measured atom-loss record,
where $t_{i,\sigma}^{(1)}$ and $t_{j}^{(2)}$ are the times when one- and two-body losses occur, respectively.
Its mean value is $\langle J\rangle=\langle I_1\rangle+\langle I_2\rangle$. The covariance of the coarse-grained record can then be written as $\left\langle\delta J(t)\delta J(0)\right\rangle=(\langle I_1\rangle+2\langle I_2\rangle)\delta(t)+\mathcal{D}(t)+\mathcal{B}(t)$,
where $\delta J(t)=J(t)-\langle J\rangle$.
The measured noise spectrum is then defined by
\begin{align}
    S_{\rm meas}(\omega)=\int_{-\infty}^{\infty}dt e^{i\omega t}\langle \delta J(t)\delta J(0)\rangle.
\end{align}
We have $S_{\rm meas}(\omega\rightarrow0)=S_N$, and 
\begin{align}\label{Smeas}
    S_{\rm meas}(\omega)\simeq \langle I_1\rangle+2\langle I_2\rangle+\mathcal{D}(\omega)+\mathcal{B}(\omega).
\end{align} 
This measured noise includes the Poissonian contribution, the connected loss-rate correlation and the jump backaction.

We now focus on a regime where the photo-induced two-body loss dominates over the one-body channel, where the covariance of loss record becomes $\left\langle\delta J(t)\delta J(0)\right\rangle=2\langle I_2\rangle\delta(t)+\mathcal{D}_2(t)+\mathcal{B}_2(t)$, where $\mathcal{B}_2(t)$ describes the quantum-jump backaction for a two-body loss channel. Importantly, this backaction contribution can be determined independently from the transient mean loss rate $\bar{I}_2(t)$ (see Appendix~\ref{appendixB}):
\begin{align}
    \mathcal{B}_2(t)=2\frac{d\bar{I}_2(t)}{dt}+O(\gamma^3_2).
\end{align}
The measured noise spectrum can be experimentally obtained by repeating destructive measurements for atom gases with different hold times. For each hold time $\tau$, the gas is
prepared under identical initial conditions, the weak two-body loss is applied, and the final atom number is measured destructively. Repeating the experiment yields the distribution of the accumulated atom loss $\Delta N(\tau)=N(0)-N(\tau)$, and its variance $\operatorname{Var}[\Delta N(\tau)] =\langle[\Delta N(\tau)-\langle \Delta N(\tau)\rangle]^2\rangle$. For a stationary loss process with finite $\tau$,
\begin{align}
    \operatorname{Var}[\Delta N(\tau)]
    =
    2\langle \Delta N(\tau)\rangle+2
    \int_0^\tau ds\,
    (\tau-s)[\mathcal{D}_2(s)+\mathcal{B}_2(s)].
\end{align}
The connected loss-rate correlation in time domain can, in principle, be reconstructed by measuring at different hold times and subtracting the leading Poissonian background:
\begin{align}\label{eq:measured-correlation}
    \mathcal{D}_2(\tau)
    =\frac{1}{2}\frac{d^2}{d\tau^2}
    &\big\{\operatorname{Var}[\Delta N(\tau)]
        -2\langle \Delta N(\tau)\rangle \big\}\nonumber\\
    &- 2\frac{d^2\langle \Delta N(\tau)\rangle}{d\tau^2}.
\end{align}
It is worth noting that while Eqs.~(\ref{eq:contact-def})--(\ref{eq:viscosity-noise}) concern the connected equilibrium contact correlator, Eqs.~(\ref{eq:loss-record})--(\ref{eq:measured-correlation}) connect this correlator to the statistics of detected losses in the weak-loss, quasi-stationary regime relevant to the experimental protocol considered here.
Such a protocol should be feasible in systems with photo-induced two-body loss, where the dissipation strength is tunable and sufficiently weak that the gas remains close to thermal equilibrium during the measurement.

An important simplification occurs at second order of $z$ in the virial expansion. 
In the $O(z^2)$ sector, a two-body loss event removes only the spin-singlet ($\uparrow\downarrow$) pair, so that no second distinct two-body loss event can occur within the same connected two-particle cluster. Consequently, the regular part of the counting covariance vanishes at this order, where the quantum-jump backaction exactly cancels the equilibrium loss-rate correlation, $\mathcal{B}_2^{(2)}(t)=-\mathcal{D}_2^{(2)}(t)$ (see Appendix~\ref{appendixB}),
where $\mathcal{B}_2^{(2)}(t)$ and $\mathcal{D}_2^{(2)}(t)$ are the second-order virial expressions of the quantum jump backaction and the connected correlator, respectively.
Therefore, the second-order virial contribution calculated in Eqs.~(\ref{eq:virial-zeta})–(\ref{eq:D-unitary}) does not appear in the regular counting noise itself, but is instead encoded entirely in the transient of the mean loss rate:
\begin{align}
    \mathcal{D}_2^{(2)}(t)=-2\frac{d\bar{I}_2^{(2)}(t)}{dt},
\end{align}
where $\bar{I}_2^{(2)}(t)$ is the second-order virial expression of the transient mean atom-loss rate.

To facilitate comparison with experiments that record the atom number as a function of time, we also consider the time-domain loss-rate correlation spectrum $\mathcal{D}_2(t)$, which can be obtained by inverse Fourier transformation of the frequency-domain spectrum in Eq.~(\ref{eq:virial-noise}). Since $\mathcal{D}_2$ is the symmetrized correlation function, we have $\mathcal{D}_2(-\omega)=\mathcal{D}_2(\omega)$. Therefore for positive $\omega$, we can write the real part of $\mathcal{D}_2(t)$ as
\begin{align}
\label{eq:redt}
    \operatorname{Re}\mathcal{D}_2(t)=\int_0^\infty\frac{d\omega}{\pi}\mathcal{D}_2(\omega)\cos(\omega t).
\end{align}

Fig.~\ref{fig:noise_tdep} shows the results of $\operatorname{Re}\mathcal{D}_2(t)$ as a function of $tT$, where the noise power decays monotonically in BCS regime and unitarity. On the BEC side, however, the bound state threshold at $\omega=E_{\rm b}$ introduces an oscillatory component, resulting in a sign change of $\operatorname{Re}\mathcal{D}_2(t)$. Because the virial spectrum has the high-frequency tail $\mathcal{D}_2(\omega)\propto\omega^{-1/2}$, leading to the short-time behavior $\operatorname{Re}\mathcal{D}_2(t)\propto t^{-1/2}$,
we use a frequency cutoff $w(\omega)=e^{-(\omega/\omega_{\rm c})^2}$ to regularize the divergence at $t=0$.

A possible experimental complication is the coexistence of multiple loss channels such as one-body and two-body dissipations. This can be resolved by the photon-induced two-body loss~\cite{PhysRevA.109.063330,PhysRevLett.132.263401}, which provides a precise and controllable source of two-body loss in a two-component Fermi gas. This allows the realization of a sufficiently large two-body loss where the effect of one-body loss can be neglected.
To ensure the dominance of two-body loss, an experimental diagnostic is the Fano factor, defined by
\begin{align}
    F(\omega\rightarrow 0)=\frac{S_N}{\langle I\rangle}=\frac{S_{\rm meas}(\omega \rightarrow 0)}{\langle I\rangle}.
\end{align}
According to Eq.~(\ref{Smeas}), the leading Poissonian contribution of order $O(\gamma)$ gives $F\rightarrow1$ for the one-body loss channel and $F\rightarrow2$ for the two-body loss channel. Since both $\mathcal{D}$ and $\mathcal{B}$ are $O(\gamma^2)$, in weak loss limit (i.e., $\gamma_{1,2}\rightarrow0$), their contribution to $F$ can be neglected in comparison with the Poissonian one. In this regard, one can identify the dominant loss channel by determining the Fano factor.

\section{Summary}\label{Sec4}
In this work, we have investigated the statistical correlation of two-body loss current recently utilized for extracting Tan's contact in ultracold atom experiments. 
Our work proposes a measurement method for the dynamical bulk viscosity $\zeta(\omega)$ through the statistics of detected losses, namely the loss-counting covariance combined with the transient mean loss rate.
We have shown that the loss-rate correlation can be reconstructed from the measured loss-counting covariance once the Poissonian self-correlation and the quantum-jump backaction, fixed by the transient mean loss rate, are subtracted.
The two-body loss current is proportional to the integrated contact density, whose equilibrium fluctuations are linked to the retarded correlator entering the Kubo formula for $\zeta(\omega)$. Based on the second-order quantum virial expansion in terms of fugacity, we have calculated the connected two-body loss-rate correlation in both frequency and time domains and presented its relation to $\zeta(\omega)$. Our results identify a realistic route to measuring the dynamical bulk viscosity through frequency-resolved loss spectroscopy.

For future perspective, it is worthwhile to extend our analysis to the case with three-body loss~\cite{PhysRevLett.112.110402} and different spatial dimensions~\cite{PhysRevLett.129.200402,PhysRevA.106.063317}.

\begin{acknowledgments}
The authors thank Shizhong Zhang and Hongchao Li for useful discussions. 
T.~Z. acknowledges support from HK GRF (Grant No.~17306024), CRF (Grants No.~C6009-20G, No.~C7012-21G, No.~C4050-23GF), CRS-HKU701/24 and a RGC Fellowship Award No.~HKU~RFS2223-7S03.
H.~T. acknowledges JSPS KAKENHI for Grants 
(Nos.~JP22K13981, JP23K22429, and JP26K07063).
\end{acknowledgments}

\appendix

\section{One-body Loss Fluctuation and Isothermal Compressibility}\label{appendixA}

The mean particle number is given by
\begin{align}
    \langle N\rangle=\frac{1}{Z}\operatorname{Tr}\big[N e^{-\beta(H-\mu N)}\big],
\end{align}
where $Z=\operatorname{Tr}e^{-\beta(H-\mu N)}$ is the partition function. We notice that
\begin{align}
    \langle N \rangle=\frac{1}{\beta}\frac{\partial \ln Z}{\partial \mu}\ \ , \qquad \frac{\partial\langle N\rangle}{\partial \mu}=\frac{1}{\beta}\frac{\partial^2\ln Z}{\partial\mu^2}.
\end{align}
Using the relation
\begin{align}
    \frac{\partial^2\ln Z}{\partial\mu^2}=\frac{1}{Z}\frac{\partial^2Z}{\partial\mu^2}-\frac{1}{Z^2}\bigg(\frac{\partial Z}{\partial \mu}\bigg)^2
\end{align}
and
\begin{align}
    \frac{\partial^2Z}{\partial\mu^2}=\beta^2\operatorname{Tr}\big[N^2 e^{-\beta(H-\mu N)}\big]=\beta^2Z\langle N^2\rangle,
\end{align}
we obtain
\begin{align}
    \frac{\partial\langle N\rangle}{\partial\mu}=\beta\big(\langle N^2 \rangle-\langle N\rangle^2\big)=\beta\langle(\delta N)^2\rangle.
\end{align}
Therefore we have
\begin{align}
    \langle(\delta N)^2\rangle=T\left(\frac{\partial\langle N\rangle}{\partial\mu}\right)_T
\end{align}
and consequently:
\begin{align}
    \mathcal{D}_{1}=\gamma^2_1\frac{T\langle N\rangle^2}{V}\kappa.
\end{align}

\section{Quantum-jump backaction and counting statistics}\label{appendixB}

\subsection{Exact loss-counting covariance}

We derive here the relation between the detected loss-counting covariance and the equilibrium loss-rate correlator. We focus on pure two-body loss channel characterized by $L_2(\bm{r})$ in Eq.~\eqref{eq:2}.
We define the integrated jump-rate operator
\begin{align}
    R=\int d^3\bm{r}L^\dagger_2(\bm{r})L_2(\bm{r}).
\end{align}
Since each jump removes 2 atoms, we have $I_2=2R$. Defining the superoperators
\begin{align}
    \mathcal{J}_2\rho=\int d^3\bm{r}L_2(\bm{r})\rho L_2^\dagger(\bm{r}),
\end{align}
and 
\begin{align}
    \mathcal{L}_2\rho=\mathcal{J}_2\rho-\frac{1}{2}\{R,\rho\},
\end{align}
we split the Liouvillian into its lossless and dissipative parts,
\begin{align}
    \mathcal{L}=\mathcal{L}_0+\mathcal{L}_2,\ \ 
    \mathcal{L}_0\rho=-i[H,\rho].
\end{align}
Let $\rho_n(t)$ denote the density matrix conditioned on \(n\) two-body loss events. It obeys
\begin{align}
    \frac{d\rho_n}{dt}=(\mathcal{L}-\mathcal{J}_2)\rho_n+\mathcal{J}_2\rho_{n-1}.
\end{align}
For the measured atom loss $N_{\rm loss}=2n$, introducing
\begin{align}
    \rho_\chi(t)=\sum_{n=0}^{\infty} e^{i 2 n \chi} \rho_n(t),
\end{align}
we obtain the tilted master equation
\begin{align}\label{Lchi}
    \frac{d \rho_\chi}{d t}=\mathcal{L}_\chi \rho_\chi, \quad \mathcal{L}_\chi=\mathcal{L}+\left(e^{i 2\chi}-1\right) \mathcal{J}_2 .
\end{align}
Here $\chi$ is the counting field used in full counting statistics. Derivatives with respect to $i\chi$ therefore insert $2\mathcal{J}_2$.

According to Eq.~(\ref{Lchi}), for two distinct times $t_2>t_1$ we have
the two-time correlation of $J(t)$ in Eq.~\eqref{eq:loss-record} as
\begin{align}
\left\langle J\left(t_2\right) J\left(t_1\right)\right\rangle  =4 \operatorname{Tr}\left[Re^{\mathcal{L}\left(t_2-t_1\right)} \mathcal{J}_2 \rho\left(t_1\right)\right],
\end{align}
where we use the identity $\operatorname{Tr}[\mathcal{J}_2 X]=\operatorname{Tr}[RX]$ for an arbitary $X$. The same-event identity $dN^2=dN$ for the counter increment $dN$ supplies the delta function term. Therefore for $t_2\geq t_1$, the exact nonstationary covariance is given by
\begin{align}
C_{J J}\left(t_2, t_1\right) \equiv &\left\langle\delta J\left(t_2\right) \delta J\left(t_1\right)\right\rangle \nonumber\\
= & 4 r\left(t_1\right) \delta\left(t_2-t_1\right)\nonumber\\ 
+&4\left\{\operatorname{Tr}\left[R{ e}^{\mathcal{L}\left(t_2-t_1\right)} \mathcal{J}_2 \rho\left(t_1\right)\right]-r\left(t_2\right) r\left(t_1\right)\right\},
\end{align}
where the instantaneous event rate is defined as $r(t)=\operatorname{Tr}[\mathcal{J}_2\rho(t)]=\operatorname{Tr}[R\rho(t)]$. If the slow evolution is frozen inside a quasi-stationary measurement window, the covariance depends only on the time difference $t$: 
\begin{align}
C_{J J}(t)= & 4 r_{\mathrm{qs}} \delta(t)
+4\left\{\operatorname{Tr}\left[R e^{\mathcal{L}|t|} \mathcal{J}_2 \rho_{\mathrm{qs}}\right]-r_{\mathrm{qs}}^2\right\}\nonumber\\
=&2\langle I_2\rangle\delta(t)
+C_{JJ}^{\rm reg}(t),
\end{align}
where we have $4r_{\mathrm{qs}}=2\langle I_2 \rangle$. The first term gives the Poissonian fluctuation of the loss current, while the second term is the regular, distinct-event part.

\subsection{Weak-loss expansion}

In the weak-loss regime, $R$, $\mathcal{J}_2$, and $\mathcal{L}_2$ are all $O(\gamma_2).$ Defining $\rho_0=Z^{-1}e^{-\beta(H-\mu N)}$ and $r_0=\operatorname{Tr}[R\rho_0]$, for times of order the intrinsic correlation time, 
we may replace the dissipative propagator in the regular two-jump term by the lossless propagator:
\begin{align}\label{Creg}
    C_{J J}^{\mathrm{reg}}(t)=4\left\{\operatorname{Tr}\left[Re^{\mathcal{L}_0 t} \mathcal{J}_2 \rho_0\right]-r_0^2\right\}+O\left(\gamma_2^3\right) .
\end{align}
We introduce the anticommutator insertion
\begin{align}
    \mathcal{A}_R\rho=\frac{1}{2}\{R,\rho\}.
\end{align}
The equilibrium loss-rate correlation $\mathcal{D}_2(t)$ can then be written as 
\begin{align}
    \mathcal{D}_2(t)=4\left\{\operatorname{Tr}\left[R e^{\mathcal{L}_0 t} \mathcal{A}_R \rho_0\right]-r_0^2\right\}.
\end{align}
Since $\mathcal{J}_2=\mathcal{A}_R+\mathcal{L}_2$, we obtain the central formula of the regular record correlation:
\begin{align}\label{eq:c-D-B}
    C_{JJ}^{\rm reg}(t)=\mathcal{D}_2(t)+\mathcal{B}_2(t)+O(\gamma^3_2),
\end{align}
where the quantum-jump backaction term reads
\begin{align}\label{backaction}
    \mathcal{B}_2(t)=4\operatorname{Tr}[Re^{\mathcal{L}_0t}\mathcal{L}_2\rho_0].
\end{align}
Therefore, in the weak-loss regime, the covariance of the measured atom loss can be written as
\begin{align}
    \left\langle\delta J\left(t_2\right) \delta J\left(t_1\right)\right\rangle\simeq2\langle I_2\rangle\delta(t)+\mathcal{D}_2(t)+\mathcal{B}_2(t),
\end{align}
where both $\mathcal{D}_2$ and $\mathcal{B}_2$ are generally $O(\gamma_2^2)$.

\subsection{Determination of $\mathcal{B}_2$ from mean loss current}

Although loss noise alone does not give the contact correlation function, the backaction contribution can be obtained from the mean response to switching on the loss. We start from the equilibrium density matrix $\rho_0$ and switch on the loss at $t=0$:
\begin{align}
    \rho_{\mathrm{on}}(t)=e^{\left(\mathcal{L}_0+\mathcal{L}_2\right) t} \rho_0 .
\end{align}
To first order in the dissipator,
\begin{align}
    \rho_{\mathrm{on}}(t)=\rho_0+\int_0^t d s e^{\mathcal{L}_0(t-s)} \mathcal{L}_2 \rho_0+O\left(\gamma_2^2\right).
\end{align}
The transient mean atom-loss rate $\bar{I}_2(t)$ can thus be given by
\begin{align}
\bar{I}_2(t) & \equiv 2 \operatorname{Tr}\left[R \rho_{\mathrm{on}}(t)\right]\nonumber \\
& =2r_0+2 \int_0^t d s \operatorname{Tr}\left[R e^{\mathcal{L}_0 s} \mathcal{L}_2 \rho_0\right]+O\left(\gamma_2^3\right).
\end{align}
Comparison with Eq.~(\ref{backaction}) gives
\begin{align}
    \mathcal{B}_2(t)=2\frac{\bar{I}_2(t)}{dt}+O(\gamma_2^3).
\end{align}
Therefore, in the weak loss regime we have
\begin{align}
    \mathcal{D}_2(t)\simeq C_{JJ}^{\rm reg}(t)-2\frac{d\bar{I}_2(t)}{dt}.
\end{align}

\subsection{Cancellation at second virial order}

We finally examine the relation between the equilibrium loss-rate correlation and the measured counting covariance within the same second-order virial expansion used in Sec. III. We write the equilibrium density matrix as
\begin{align}
    \rho_0=\frac{1}{Z}\sum_{N=0}^{\infty}z^N e^{-\beta H_N} P_N,
\end{align}
where $P_N$ projects onto the $N$-particle sector, and $H_N$ represents the Hamiltonian restricted to the sector containing exactly $N$ particles. We note that the two-body jump operator lowers the particle number by two,
\begin{align}
    L_2(\bm{r})P_N=P_{N-2}L_2(\bm{r}),
\end{align}
whereas the rate operator $R$ preserves particle number and vanishes in the $N=0$ and $N=1$ sectors: $[R,N]=0$, $RP_0=RP_1=0$. Then consider the ordered two-jump contribution in Eq.~(\ref{Creg}),
\begin{align}
    K(t)=\operatorname{Tr}\left[Re^{\mathcal{L}_0 t}\mathcal{J}_2\rho_0\right].
\end{align}
For $N=2$, the first two-body jump takes the system into the vacuum sector, where $R=0$. Likewise, for $N=3$, the first jump leaves a one-particle state, for which $R=0$. Therefore, neither the $N=2$ nor the $N=3$ sector contributes to $K(t)$, and the first nonvanishing contribution arises from $N=4$: $K(t)=O(z^4)$. Since $r_0=\operatorname{Tr}[R\rho_0]=O(z^2)$, the disconnected contribution $r_0^2$ is also $O(z^4)$. Therefore its second-order virial contributions vanish:
\begin{align}
    C_{JJ}^{{\rm reg}(2)}(t)=0.
\end{align}

In contrast, the anticommutator insertion $\mathcal{A}_R\rho=\{R,\rho\}/2$ preserves particle number, so the $N=2$ sector contributes directly to the equilibrium correlator $\mathcal{D}_2(t)$. Therefore $\mathcal{D}_2^{(2)}(t)$ is generally nonzero. Comparing with Eq.~(\ref{eq:c-D-B}), we obtain
\begin{align}
    \mathcal{D}_2^{(2)}(t)=-\mathcal{B}_2^{(2)}(t)=-2\frac{d\bar{I}_2^{(2)}(t)}{dt}.
\end{align}
Physically, the $O(z^2)$ term describes a single connected two-particle cluster. Its contact can fluctuate in equilibrium, giving a finite $\mathcal{D}_2^{(2)}(t)$, but once a two-body loss event occurs the pair is removed and cannot generate a second distinct loss event. Event-event correlations therefore require at least two pairs and start only from $O(z^4)$.

\subsection{Quantum-jump backaction in the one-body loss channel}

For completeness, we show that the measured loss-counting covariance in the one-body channel contains a quantum-jump backaction contribution $\mathcal{B}_1$. Applying the counting-field analysis to the one-body channel, we obtain
\begin{align}
    \mathcal{B}_1(t)=\frac{d \bar{I}_1(t)}{dt}+O(\gamma_1^3),
\end{align}
where $\bar{I}_1(t)$ is the transient mean one-body loss rate.
Since $I_1=\sum_{\sigma}\gamma_{1\sigma}N_{\sigma}$ and $N_\sigma$ is conserved by the equilibrium Hamiltonian, this backaction reflects only the slow depletion of the particle number and does not generate intrinsic dynamical correlations.
For spin-dependent one-body losses,
\begin{align}
    \mathcal{B}_1(t)=-\sum_{\sigma}\gamma_{1\sigma}^2\langle N_\sigma\rangle.
\end{align}
Therefore, the equilibrium one-body loss-rate correlation remains purely static and is connected to the isothermal compressibility, as shown in Eq.~(\ref{D1-kappa}).

\bibliography{reference.bib}

\end{document}